\documentstyle[epsfig,prl,twocolumn,aps]{revtex}

\begin{document}

\title{Reconstruction of dynamical equations for traffic flow}
\author{S. Kriso$^1$, R. Friedrich$^2$, J. Peinke$^1$ and P.
Wagner$^3$}
\address{%
$^1$ Fachbereich 8 Physik, Universit\"at Oldenburg, 26111 Oldenburg, Germany\\
$^2$Institut f\"ur Theoretische Physik und Synergetik, Universit\"at
Stuttgart, 70550 Stuttgart, Germany \\
$^3$ Deutsches Zentrum f\"ur Luft- und Raumfahrt (DLR) e.V.,
Institut f\"ur Verkehrsforschung, Rutherfordstrasse 2, 12489 Berlin, Germany
}
\date{\today}

\maketitle

\begin{abstract}
  Traffic flow data collected by an induction loop detector on the
  highway close to K\"oln-Nord are investigated with respect to their
  dynamics including the stochastic content.  In particular we present
  a new method, with which the flow dynamics can be extracted directly
  from the measured data.  As a result a Langevin equation for the
  traffic flow is obtained.  From the deterministic part of the flow
  dynamics, stable fixed points are extracted and set into relation
  with common features of the fundamental diagram.
\end{abstract}

\section{Introduction}

By increasing the number of licensed vehicles on our roads it becomes
more and more necessary to reduce appearances of traffic congestion.
Getting higher capacities of highways means looking for optimized flow
rates of cars. In order to do so, it is necessary to investigate the
complex system of traffic flow and to understand its regularities. In
a subsequent step, intelligent traffic control systems may use these
laws to influence the traffic flow and thus to increase the highway's
capacity utilization.

For our investigation a large amount of traffic flow data were
collected at the highway near K\"oln-Nord (Germany) over more than one
week. For each car crossing an induction loop detector the following
data were recorded: (i) the time, when the car crossed the
detector, with an resolution of 1 sec, (ii) the type of car (passenger car,
truck), (iii) the lane
number, (iv) the velocity, truncated to an 8-bit integer (0...255
km/h), accuracy of approximately 3 \%, (v) the length of the car in
meters (8-bit integer, not calibrated) and (vi) the distance to the
car driving ahead (an integer in the range 0\ldots 999 m, not calibrated).
The lanes are
labeled from A to C, where lane A is the right driving lane where
mostly trucks are found (because of a German law), and lanes B and C
are the fast drivers' lanes.

In the following we present a new method to derive from the measured
data dynamical equations for the traffic situation.  We start with the
common presentation of the fundamental diagram.  Next, we evaluate the
deterministic and the stochastic content of of the traffic dynamics by
means of a Langevin equation.  Finally, we give an interpretation of
the determined Langevin equation.

\section{Fundamental diagrams}

Here and in the following we calculate the car density $k$ [km$^{-1}$]
and the flux state $q$ [h$^{-1}$] of each car using the measured data:
the velocity $v$ [km/h], the length of the car $l$ [m] and the
distance $d$ [m] to the car ahead:
%

\begin{eqnarray}
  k & = & \frac{1000}{l+d} \;\; ,\\ q & = & k \cdot v\;\; .
  \label{Def}
\end{eqnarray}

To avoid an overload of the presentation we restrict the diagrams and
the following calculations to two cases.  Firstly, only the traffic of
a single lane C, secondly the cumulative traffic of all three lanes A,
B and C are considered.  Furthermore we calculate for each density
state $k$ the mean flux state

\begin{equation}
	\langle q \rangle_{k} = \frac{1}{n(k)} \sum q(k) \;\; .
	\label{eins}
\end{equation}

\begin{figure}[ht]
   \begin{center} 
      \epsfig{file=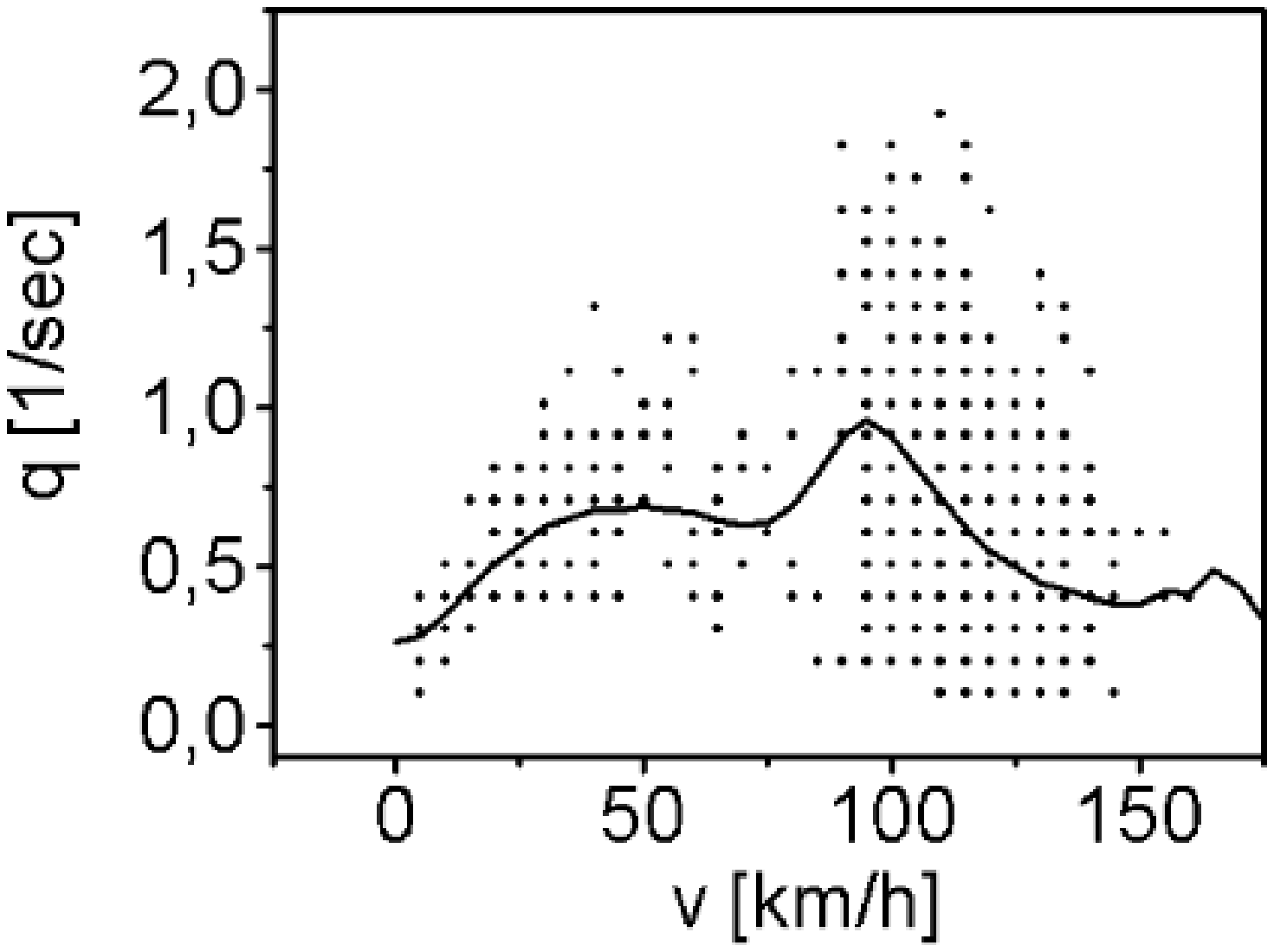, width=6.0cm}
      \epsfig{file=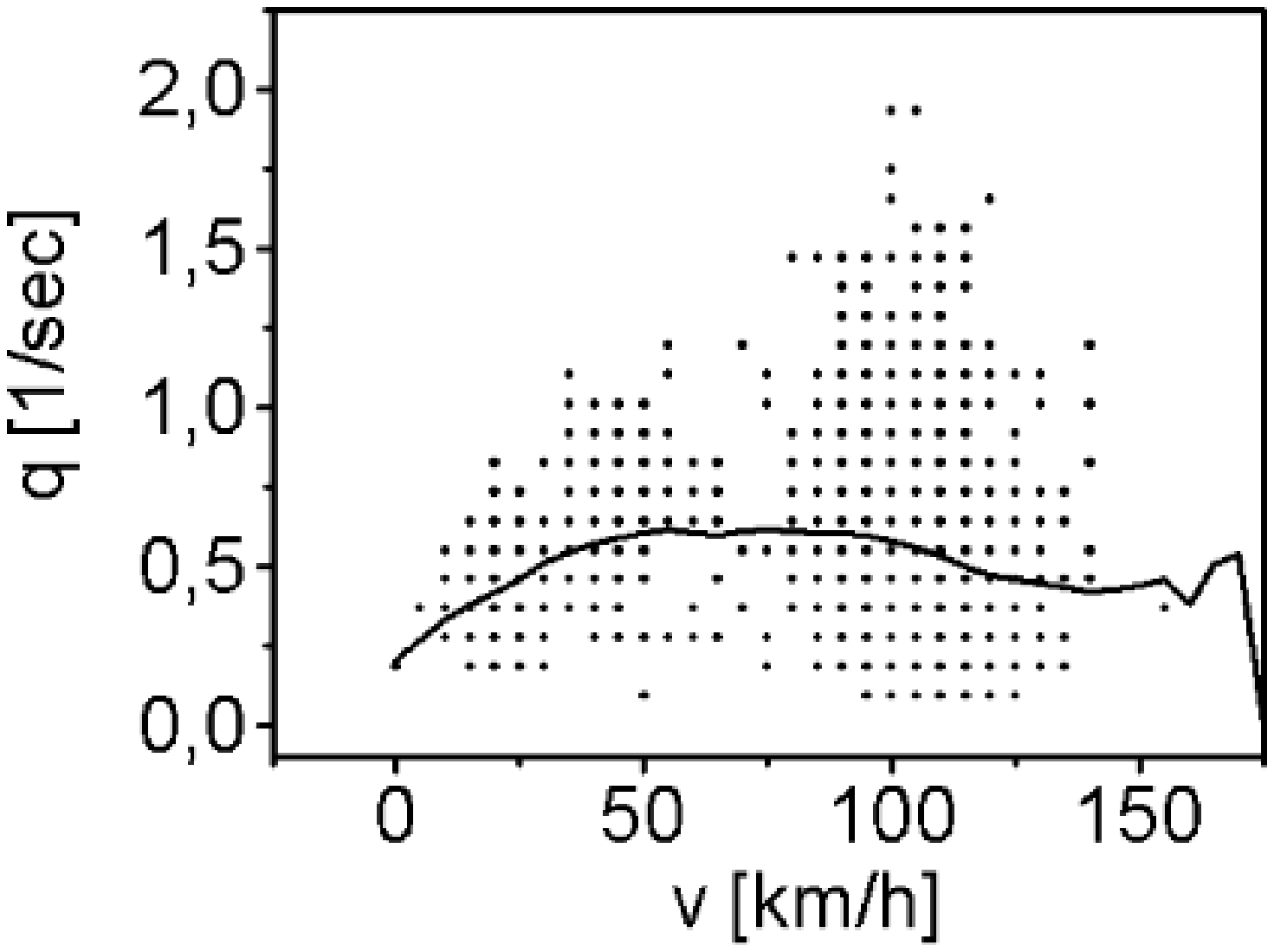, width=6.0cm}
   \end{center}
   \caption{Fundamental diagram of (a) lane C and of (b) all three
   lanes A,B, and C. The dots represent the measured traffic data, the
   solid line corresponds to the mean flux $\langle q \rangle_k$.}
\label{FundDia}
\end{figure}

The obtained fundamental diagrams (q$_{i}$;k$_{i}$) and ($<$q$>$;k)
are shown in Fig.~\ref{FundDia}.  We note that for this presentation
no significant difference in the traffic dynamics of one lane and the
cumulative dynamics of three lanes can be detected. In both diagrams
we find a maximum flux in free traffic flow of $q_{\rm max} = 8000$
cars/h, according to the results of \cite{Kerner}.  For the flux out
of traffic jams we find in both cases $q_{\rm out} = 5000 $ cars/h.
So we have a ratio of $\eta := q^{\rm free}_{\rm max}/q_{\rm out} =
1.6$, which meets the value $\eta \approx 1.5$ found in \cite{Kerner}
quite well.

\section{Langevin equation for the traffic flow}

In order to grasp the underlying dynamics of the traffic flow (one
lane and all three lanes), we utilize a new method to analyze the
traffic data more extensively.
In particular, the iterative dynamics of the traffic state ${\mathbf
x}_{N}$, given by the velocity ${\mathbf v}_N$ and the flux ${\mathbf
q}_N$ of the $N$-th car, as a function of traffic state ${\mathbf
x}_{N-1}$ of the $N-1$-th car is investigated. Note, that other state
variables could have been choosen as well.

For the iterative dynamics of the traffic state ${\bf x}_{N}$ (which
may be taken in a generalized case as a $r$-dimensional variable) we
propose a description by a stationary Langevin equation, taking into
account a combination of deterministic and random (noisy) forces cf.
\cite{Risken}:
\begin{equation}
	x_{i,N+1} = x_{i,N} + h_{i}({\bf x}_{N}) + \sum_{j=1}^{r}
	g_{ij}({\bf x}_{N}) \cdot 	\Gamma_{j,N} \;\; ,
	\label{Langevin}
\end{equation}
where the indices $i$ and $j$ denote components of the
multidimensional variables, $N$ the time, and $\Gamma_{j,N}$ are $r
\times N$ independent Gaussian noise variables with zero mean and with
variance 2, i.e.,
\begin{equation}
	\langle \Gamma_{j,N} \rangle = 0 \; , \;\;
	\langle \Gamma_{j,M} \cdot \Gamma_{k,N} \rangle =
	2 \delta_{j,k}\delta_{M,N} \;\; .
\end{equation}

The central part of our following work is that it is possible to
determine the functions $h_{i}$ and $g_{ij}$ directly from empirical
data. Taking (\ref{Langevin}) as the Ito presentation of the
stochastic process the following relation to the Kramers-Moyal
coefficients can be given,
\begin{eqnarray}
   	D_{i}^{(1)}({\bf x}) & = & h_{i}({\bf x}) \\
	D_{ij}^{(2)}({\bf x}) & = &\sum_{k=1}^{r}g_{ik}({\bf
x})g_{jk}(		{\bf x}) \;\; ,
	\label{Dxh}
\end{eqnarray}
where $D^{(1)}$ and $D^{(2)}$ are called drift and diffusion
coefficient.
These coefficients can be evaluated by the conditional moments
\begin{eqnarray}
	 \label{D1} D_{i}^{(1)}({\bf x}) = \langle x_{i,N+1}-x_{i}
\rangle \Big|_{{\bf
	x}_{N}={\bf x}} \;\; , \\
	D_{ij}^{(2)}({\bf x}) =
\langle (x_{i,N+1}-x_{i})(x_{j,N+1}-x_{j}) \rangle \Big|_{{\bf
	x}_{N}={\bf x}} \;\;  .
	\label{DefD}
\end{eqnarray}

Recently it has been shown that with the analogous definition of
these Kramers-Moyal coefficients it is possible to reconstruct from
time continuous dynamics the underling stochastic differential
equation \cite{us}.

Before presenting our results on the dynamics we want to comment
on the validity of this ansatz to describe the traffic flow by the
Langevin equation (\ref{Langevin}). This ansatz implies that the
dynamics is in the class of Markovian processes, i.e. the system does
not have a memory. This can be tested by conditional probabilities
\begin{equation}
	p(\bf{x}_{N}|\bf{x}_{N-1}, \ldots, \bf{x}_{N-m}) =
	p(\bf{x}_{N}|\bf{x}_{N-1})
\end{equation}
or by the necessary condition of the Chapman-Kolmogorov equation
\begin{equation}
		p(\bf{x}_{N}|\bf{x}_{N-r}) = \sum_{\bf{x}_
{N-s}} 	p(\bf{x}_{N}|\bf{x}_{N-s})
		p(\bf{x}_{N-s}|\bf{x}_{N-r}) \;\; ,
\end{equation}
where $r>s$. From our data, conditional probabilities have been
evaluated and the validity of the Chapman-Kolmogorov equation was
found for the iterative dynamics of both quantities, the velocity and
the flux. If this Markovian property holds, the inherent noise of the
dynamics ($\Gamma$) can be taken as $\delta$-correlated. It should be
noted, that even in the case where the noise is not
$\delta$-correlated, the deterministic part of the dynamics can be
reconstructed from given data (\ref{D1}), as we found by analysing
numerically generated test data, \cite{Malte}.

\begin{figure}[ht]
   \begin{center}
      \epsfig{file=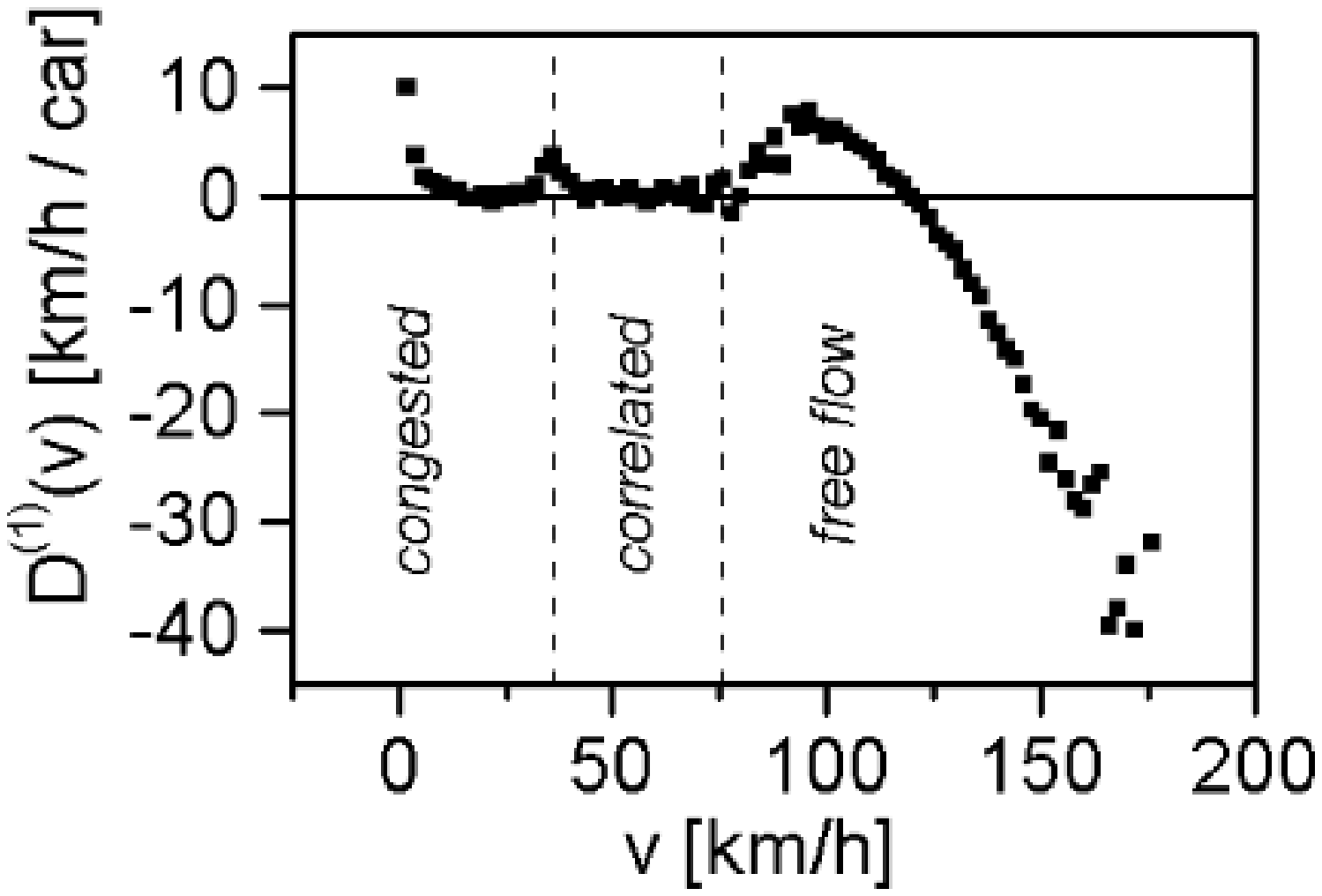, width=6.0cm}
      \epsfig{file=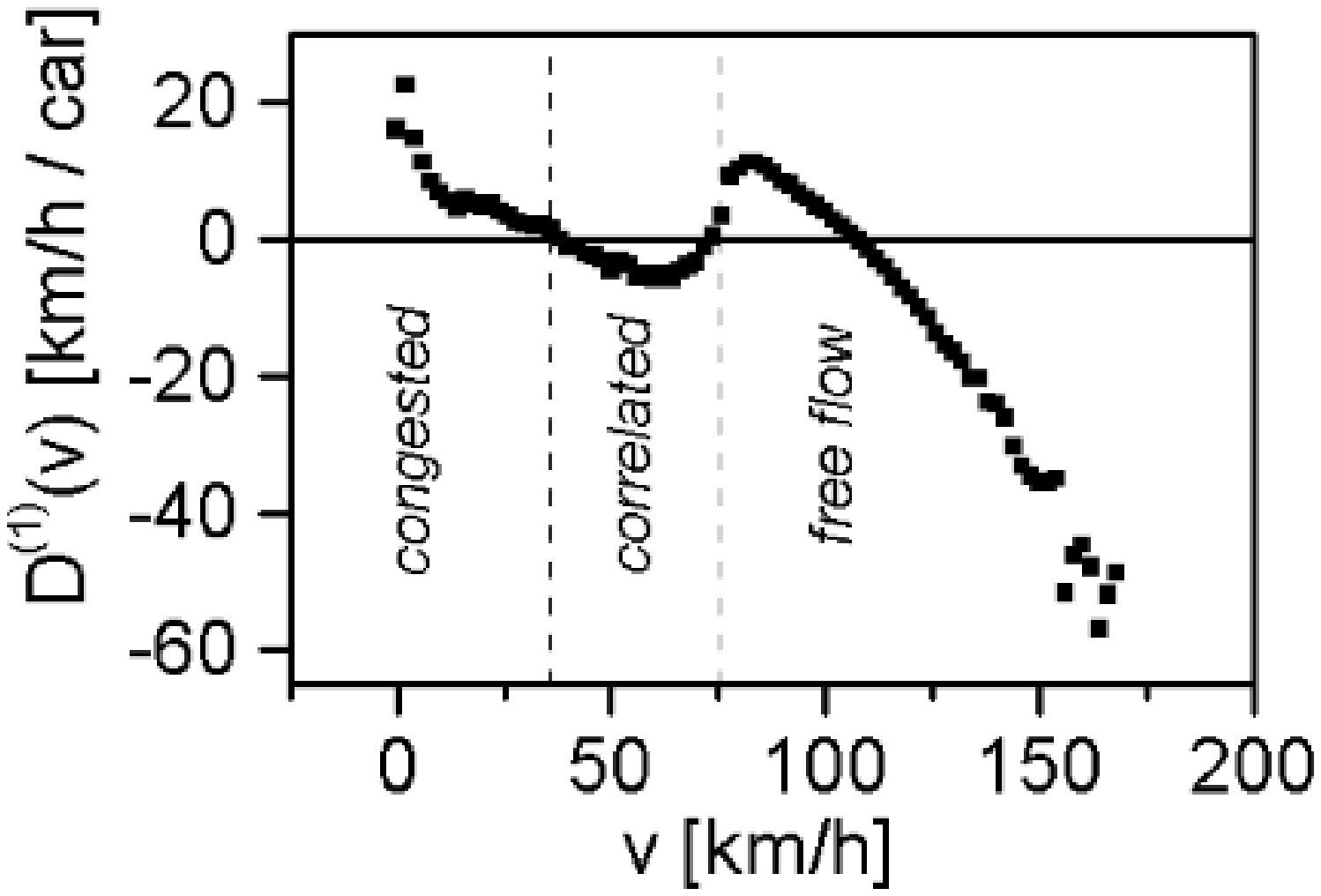, width=6.0cm}
      \epsfig{file=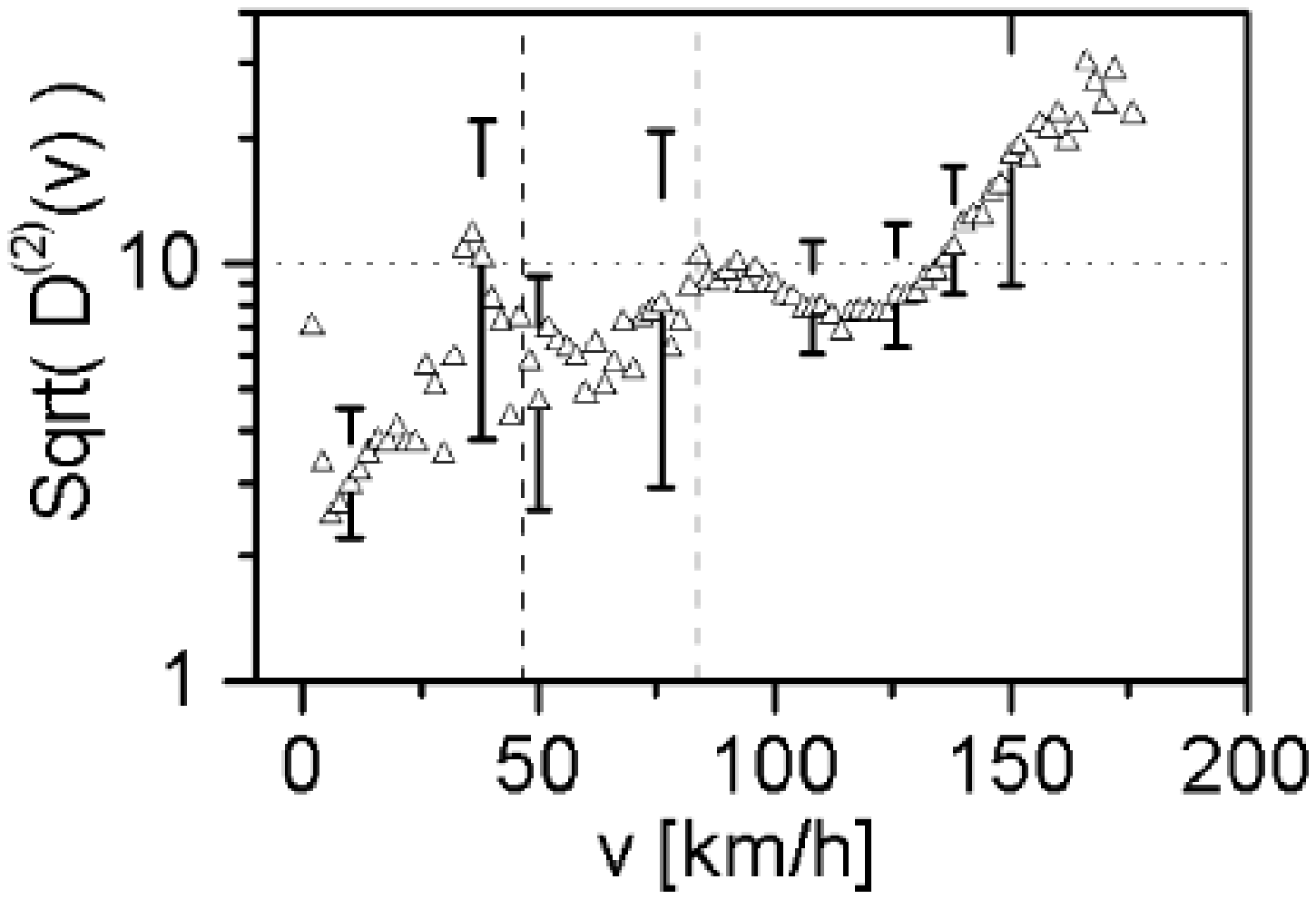, width=6.0cm}
      \epsfig{file=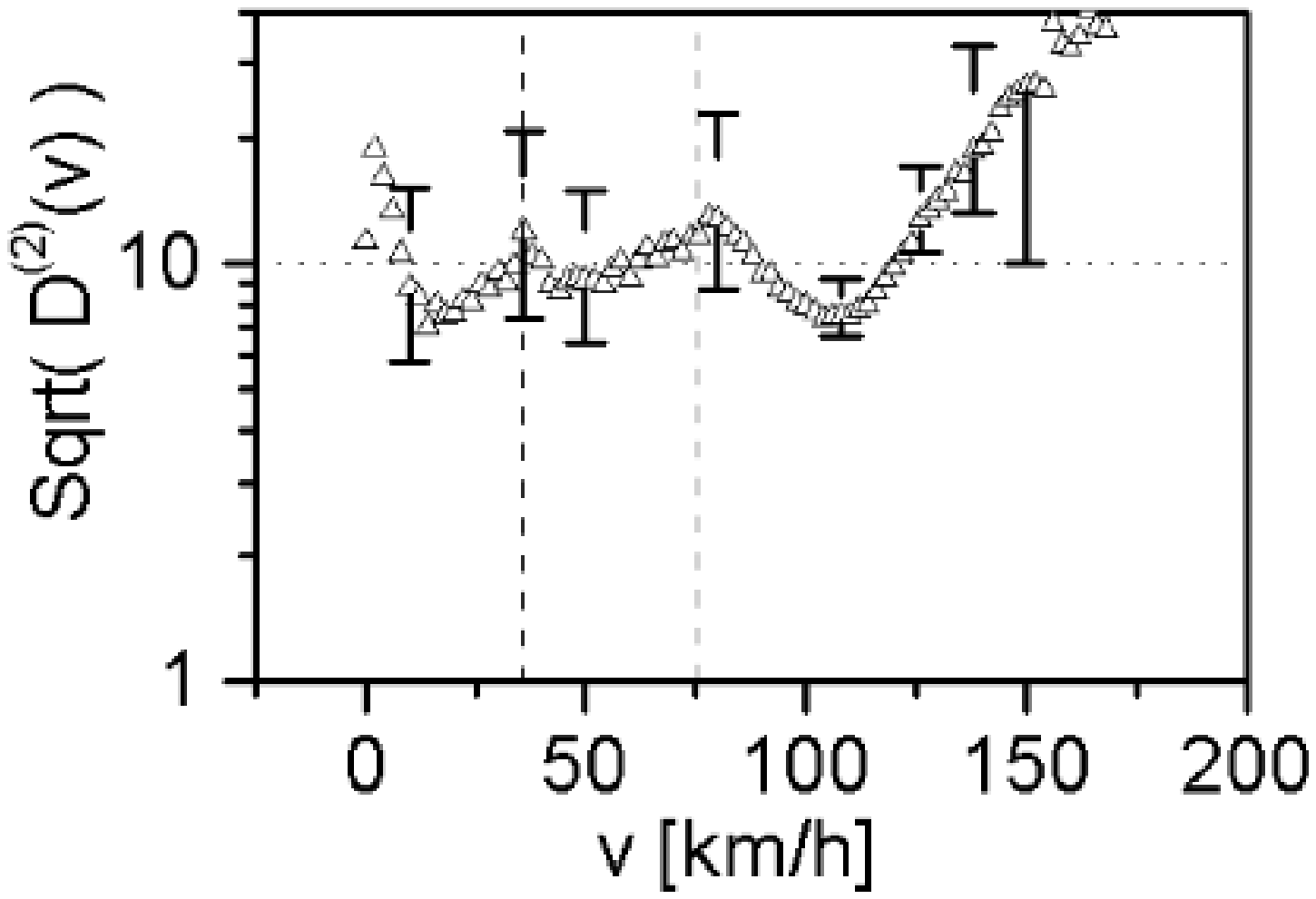, width=6.0cm}
   \end{center}
  \caption{Drift and diffusion coefficients for the traffic of one lane
  C, (a) and (c) and of
  all three lanes (b) and (d). }
  \label{Drift}
\end{figure}

\begin{figure}[ht]
   \begin{center}
      \epsfig{file=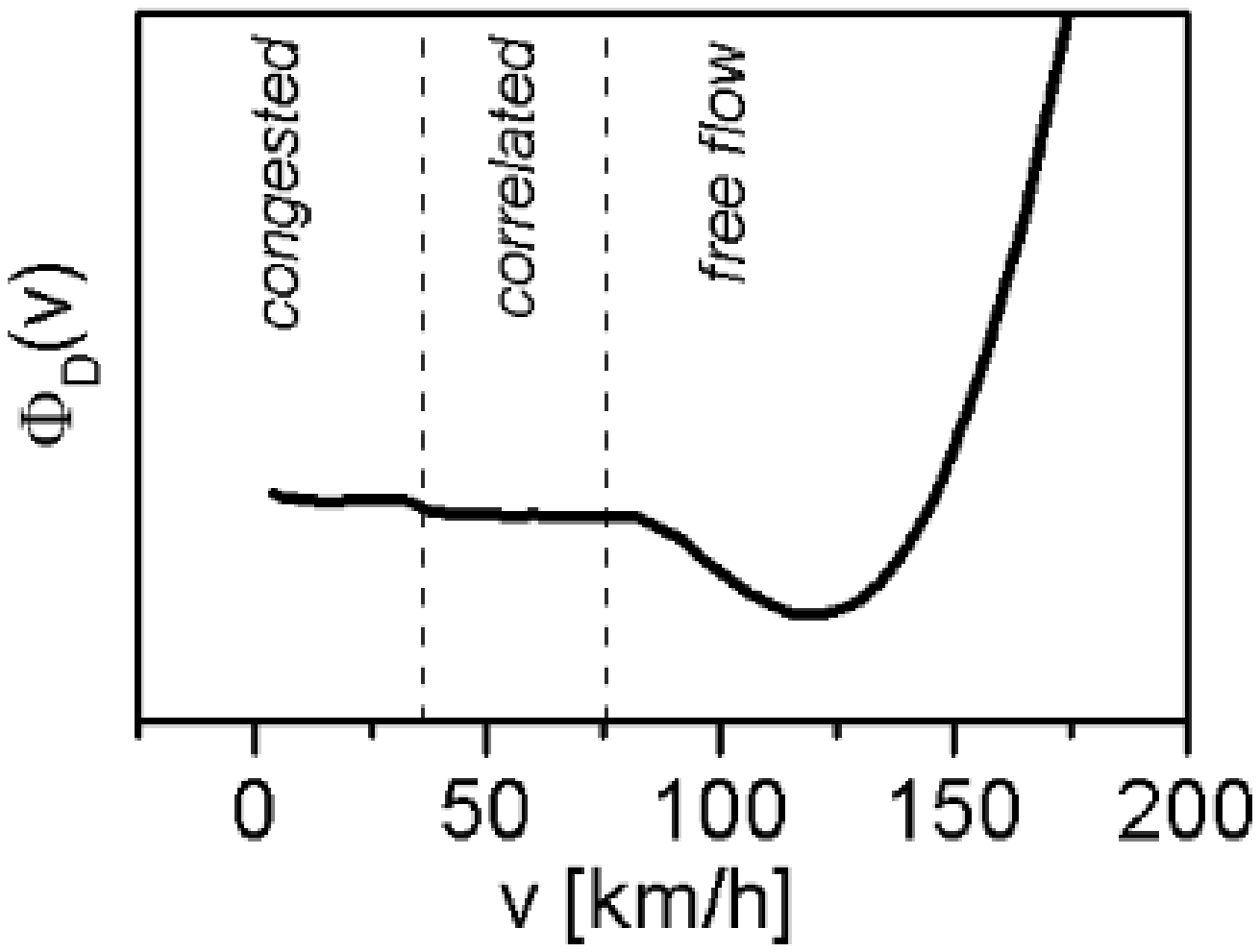, width=6.0cm}
      \epsfig{file=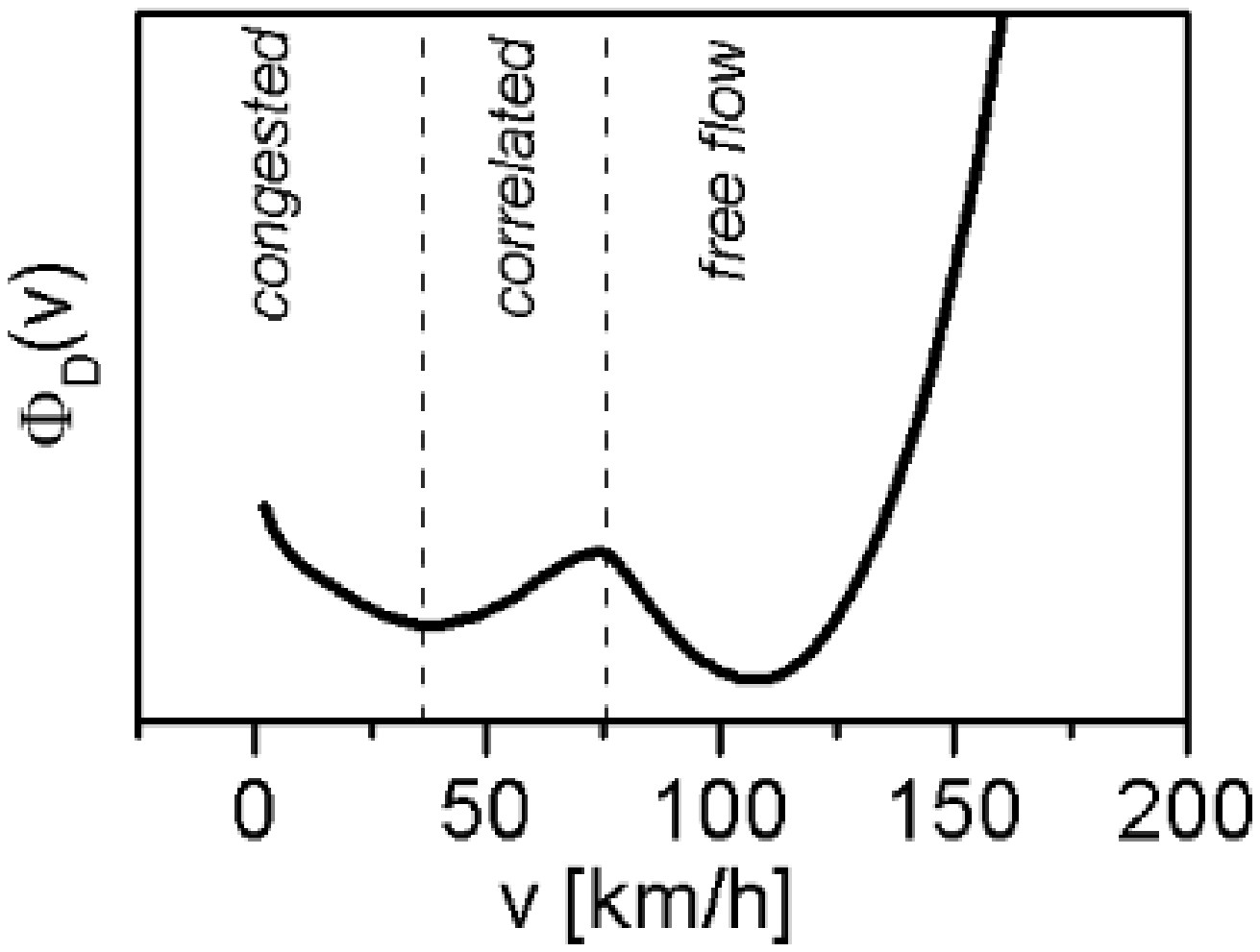, width=6.0cm}
   \end{center}
  \caption{Corresponding potentials for the deterministic dynamics
  given by the drift coefficients in Fig.~\ref{Drift}.}
  \label{Pot}
\end{figure}

As expressed by (\ref{DefD}), the knowledge of the conditional
probabilities $p(\bf{x}_N | \bf{x}_{N-1})$ provides the basis to estimate the
Kramers-Moyal coefficients from the traffic data.  First we consider
the simplified case of the onedimensional dynamics of the velocity
only.  The results for and  $D^{(2)}$(v) for the traffic of one lane and
for the cumulative traffic of all three lanes are shown in
Fig.~\ref{Drift}. The one dimensional deterministic dynamics
can also be expressed by the potential $\Phi_{D}$(v), defined as -
$\frac{\delta \Phi_{D}}{\delta v}$ = D$^{(1)}$(v). The corresponding
potentials are shown in Fig.~\ref{Pot}. From these results
three noticable velocities $v_1 \approx 37$ km/h, $v_{2} \approx $ 75
km/h and $v_{3} \approx $ 107 km/h appear, which allows to identify
the so called congested flow for $v\leq v_1$, correlated flow for
$v_{1} < v \leq v_{2}$ and the free flow for $v > v_{2}$
\cite{flowDef,flowDef1}, respectively. Note these velocities can be
defined as fixed points ($D^{(1)}(v)=0$) of the deterministic part of the
cumulative traffic dynamics (see Fig. 2b).

For the traffic dynamics of lane C we find in the congested and in
the correlated regime metastable traffic states, the deterministic
drift term $D^{(1)}$ gets zero over finite intervals. This
corresponds to the plateau structure in the potential, see Fig. 3a. A clearly
different behaviour is found for the cumulative traffic dynamics of
all three lanes, see Figs.~\ref{Drift}b and \ref{Pot}b. The different
flow regimes are seperated by two fixed points at $v_{1}$ and
$v_{2}$. The slope of these fixed points defines the stability, thus
the congested and the correlated flow regimes are separated by a
stable fixed point, whereas the correlated and the free flow regime
are separated by an instable fixed point. For the free flow, in both
cases of one lane or three lane traffic a stable fixed point is found
at $v_{3}$, correspondingly the drift potential has its local minimum.
Because of a speed limit of 100 km/h given on the inspected highway we
see a great increase of the potential for $v > v_{3}$: the faster a
car is driving, the stronger the attraction is to the potential's
minimum \cite{comment} .

To get an understanding of the real traffic dynamics grasped by these
drift coefficients or drift potentials, the additional noise has to be
taken into account. In Fig.~\ref{Drift}c and d the corresponding
magnitude of the noise are expressed by the evalutaed diffusion
coefficients $D^{(2)}$. The noise will now cause transitions between
different flow states. For the traffics dynamics of one lane, the
noise will effect larger fluctuations as it is the case for the
cumulative traffic dynamics, which has two clear minima in the
potential. A further interesting detail is that the
magnitude of $D^{(2)}$ has a minimum around the stable fixed point at
$v_{3}$. This indicates a pronounced stability of this traffic
situation.

\begin{figure}[ht]
   \begin{center}
      \epsfig{file=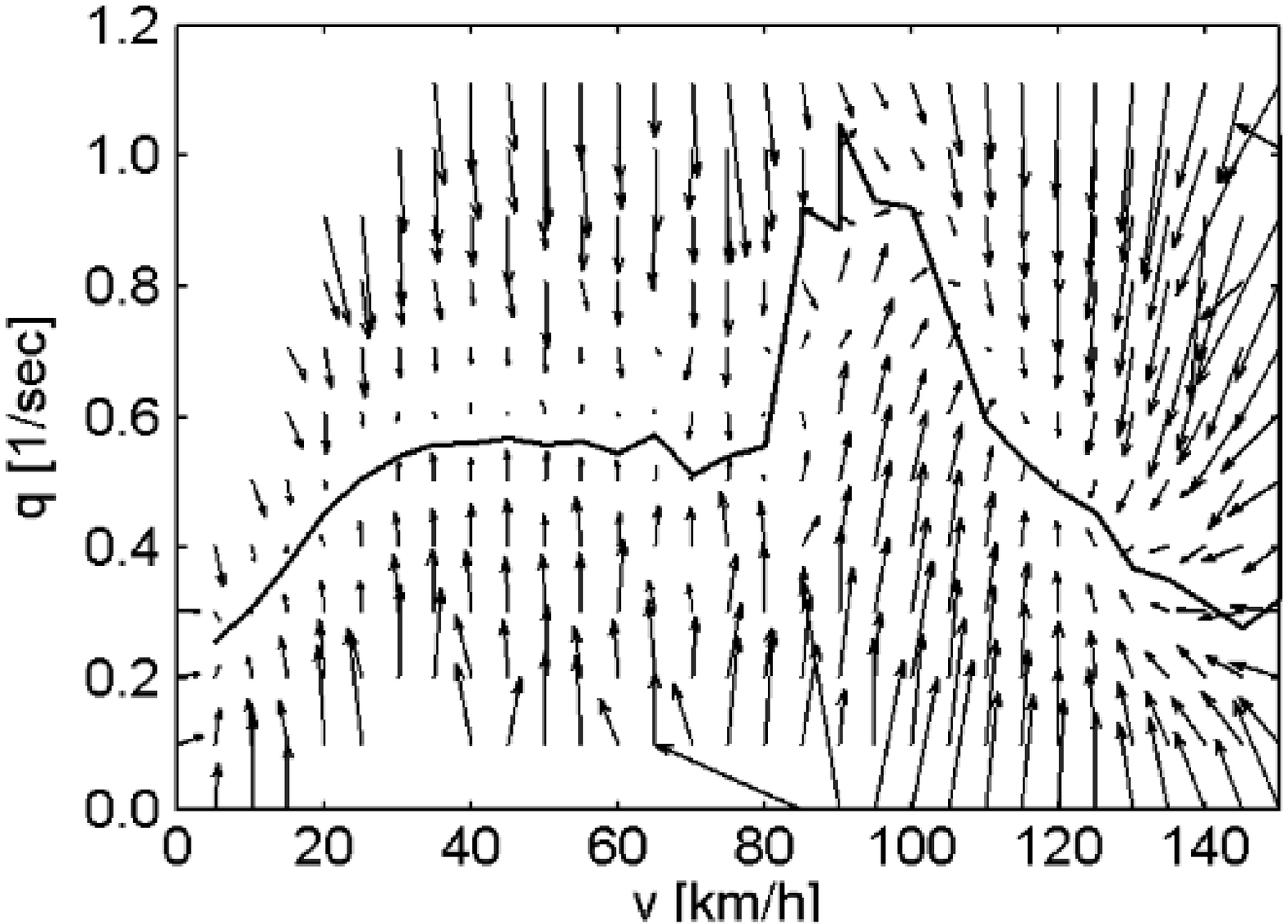, width=6.0cm}
      \epsfig{file=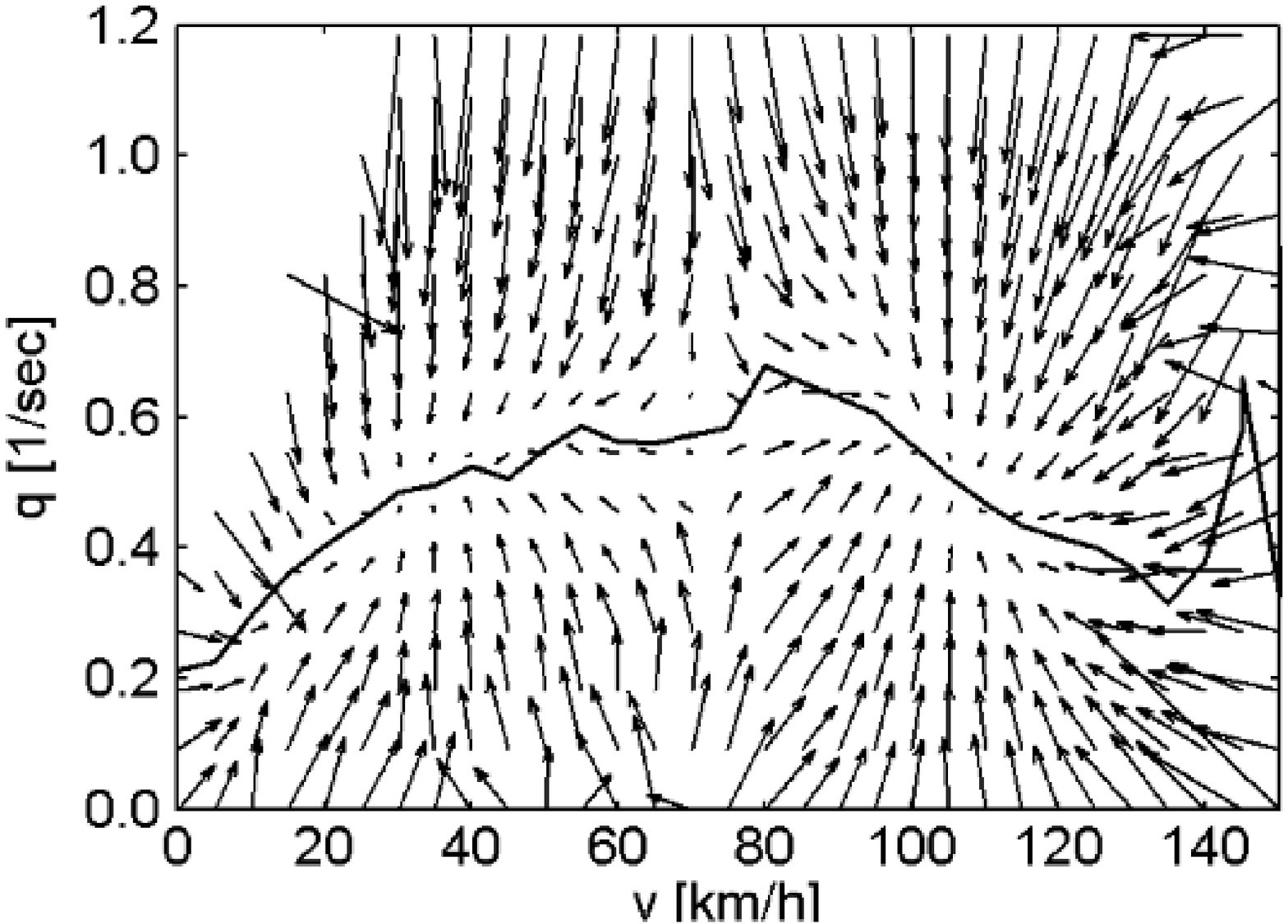, width=6.0cm}
   \end{center}
  \caption{Deterministic dynamics of the twodimensional traffic states
  $(q,v)$ given by the drift vector for (a) one lane C and (b) all three
  lanes. Bold dots indicate stable fixed points and open dots saddle
  points, respectively.}
  \label{2DDrift}
\end{figure}

Next we present the results of a higher dimensional analysis by taking
${\bf x}$ with the components $x_{1} = v$ and $x_{2} = q$. Now also
the drift coefficient $D^{(1)}$ becomes a vector depending on $v$ and
$q$, as shown in Fig.~\ref{2DDrift}. These results were obtained
by binning the velocity flux data into a $25 \times 25$ matrix,
corresponding to a binning of the velocity into intervals of $\Delta
\, v = 5$ km/h.

The solid lines in Fig.~\ref{2DDrift} show the states where we have no
drift of the flux component: $D_{q}^{(1)} = 0$. On these lines we find
only a velocity drift with a constant flux. In accordance to
Fig.~\ref{Drift}a we find in Fig.~\ref{2DDrift}a for slow velocities
mainly no velocity drift, in Fig.~\ref{2DDrift}b (like in
Fig.~\ref{Drift}b) there seem to exist stable velocity drift states at
the same velocities ($v_i \approx 37$km/h,$107$ km/h). The topology of
the instable fixed point gets now a saddle point which is attractive
for larger and smaller flux values but instable in the direction of
larger and smaller velocity values. Again clear differences of the
dynamics of on lane and three lanes is found.

\section{Discussion and Conclusion}

By the investigation of traffic flow data as an iterative stochastic
process we were able to calculate from the given data the
1-dimensional drift and diffusion coefficients and thus to find the
deterministic and stochastic part of the corresponding Langevin
equation.  We are able to find stable, metastable and unstable states
(fixed points) in the deterministic part of free, correlated and
congested traffic flow.  For a fully description of the whole dynamics
also the diffusion coefficient has to be taken into account, which
provides transition probabilities between the different (meta-)stable
states.  Without this noisy part of the traffic flow dynamics, the
stable states would never be left, i.e.\ a congestion would stay
forever if once prepared.

To see the dependency of velocity and flow from each other,
the investigations were expanded to a higher dimensional analysis.  In
this case we find the deterministic and stochastic part of the
2-dimensional Langevin equation. Now we are able to identify stable
velocity and flux states of the deterministic part.

Interestingly, the results found in this study are in agreement with
empirical investigations that have identified three phases of traffic
\cite{phastrans}, together with transitions connecting theses phases.
Especially the transition from free flow to correlated flow (or
synchronized flow in the terminology of \cite{phastrans}) has
similarities.

Finally we want to point out, that we presented here a new method to
analyse traffic data with respect to a derivation of dynamical
equations from pure data analysis. Furthermore we could show that
our method provides more insight into the traffic dynamics than the
presentations of diagrams like the fundamental diagram. A clear
difference in the dynamics of one lane and the dynamics of cumulative
three lanes was found. Our analysis provides evidence of the presence
of fixed points, which are of practical importance if a control of a
traffic should be achieved. At last one may conclude that this method
will be helpful to perform a more thorough comparison between traffic
flow models and empirical data.

Acknowledgement: Helpful discussions with Ch.~Renner, St.~L\"uck and
M.~Siefert are acknowledged. We also would like to thank the
Landschaftsverband Rheinland and the Northrhine-Westfalia Ministry for
Economy and Transport for providing the data used in this study.

%
%
%
%
%
%
%
%
%
%
%

\begin{references}
\bibitem{comment} It is commonly known, that in Germany people like
to drive about 10 percent faster than given speed limits. With this
speed there is still no punishment by the police.
\end{references}
\end{document}